\documentclass[aps,prb,twocolumn]{revtex4-1}
\usepackage{graphicx}
\usepackage{amssymb,amsfonts,amsmath}
\usepackage[colorlinks=true,citecolor=Cerulean,linkcolor=RubineRed,urlcolor=Cerulean]{hyperref}
\hypersetup{breaklinks=true}
\usepackage{graphicx}
\usepackage{color}
\usepackage[usenames,dvipsnames]{xcolor}
\usepackage{epstopdf}
\usepackage{units}

\newcommand{\Jso}{J_\mathrm{so}}

\newcommand{\Htb}{H_\mathrm{latt}}
\newcommand{\sx}{\langle S_x \rangle}
\newcommand{\sz}{\langle S_z \rangle}

\begin{document}

\title{Controlling spin without magnetic fields: the Bloch-Rashba rotator}

\author{C.E.~Creffield}
\affiliation{Departamento de F\'isica de Materiales, Universidad
Complutense de Madrid, E-28040 Madrid, Spain}

\date{\today}

\begin{abstract}
We consider the dynamics of a quantum particle held in a lattice potential,
and subjected to a time-dependent spin-orbit coupling. Tilting the
lattice causes the particle to perform Bloch oscillations, and by suitably
changing the Rashba interaction during its motion, the spin of the 
particle can be gradually rotated. Even if the Rashba coupling can only
be altered by a small amount, large spin-rotations 
can be obtained by accumulating the rotation from successive oscillations.  
We show how the time-dependence of the spin-orbit coupling can be
chosen to maximize the rotation per cycle, and thus how this method can be used
to produce a precise and controllable spin-rotator, which
we term the Bloch-Rashba rotator, 
without requiring an applied magnetic field.
\end{abstract}

\maketitle

\section{Introduction}
Spintronics \cite{spintronics} is a rapidly developing field of study,
in which information is carried by an electron's spin as well as its
charge. Spin qubits not only have the benefit of long spin coherence times,
but their lower energy scales also promise lower-power, higher-speed devices.
Their implementation, however, requires a method of manipulating the spin 
of individual electrons.
This can be done by using micromagnets
\cite{micromagnets_1,micromagnets_2,micromagnets_3,micromagnets_4,bayat},
but it is difficult to confine magnetic fields to the small volumes
occupied by the qubits and obtain the necessary level of control.
A method of addressing spin by applying local gating potentials
would thus be greatly preferable. A possible means to achieve this is
provided by spin-orbit coupling (SOC). 
This is a relativistic effect in which
an electric field is transformed into an effective magnetic field
in the rest-frame of the electron,
which then interacts with the electron's spin, 
coupling it to the particle's momentum.
In condensed matter systems,
SOC underlies the existence of topological insulators \cite{top_ins},
and provides the basis of the spin quantum Hall effect \cite{spin_hall}.

If the electric field arises from inversion asymmetry in the crystal
lattice itself, the SOC is termed Dresselhaus coupling. Alternatively, if it
arises from spatial inhomogeneity of a heterostructure interface,
it is called Rashba coupling \cite{rashba_old,rashba}. 
The Rashba effect is particularly suitable for qubit
manipulation because the magnitude of the coupling
can be tuned by electrostatic gates
\cite{nitta,liang}. An electron's spin can thus be rotated by
moving an electron in space while controlling the size of the Rashba
coupling \cite{rashba_1}. In a quantum wire, for example, a spin-flip
can be obtained \cite{wire} by allowing an electron to move a certain 
distance along the wire, where the required distance \cite{wolfgang}
is inversely related to the strength of the coupling. 

It would, however, be more convenient to be able to transport the electron back to its original location, so that having been rotated by a certain
angle it can then be used for further quantum logic operations.
This requires time-dependent control of the Rashba coupling 
\cite{oscillating_so_4,cohen,referee_prb,oscillating_so_5,referee_epjb,bednarek},
otherwise the rotation-angle obtained on the outward leg of the electron's
journey would be unwound by the return leg. In Ref. \cite{cadez_2}
a method to achieve this was proposed, where an electron
trapped in a local potential was moved along a closed
trajectory in space, while the Rashba coupling was
varied in time, to obtain the desired spin-rotation. 
An appealing aspect of this system is that
exact analytical solutions can be obtained 
\cite{toni_john,kregar,cadez}, allowing its robustness towards gate noise
\cite{gate_noise} and thermal effects \cite{donvil}
to be assessed.

In this paper we consider inducing a spin-rotation in a conceptually
similar way, but instead we use a lattice system.
This could be produced
by applying a superlattice potential to a quantum wire, or by
suitably gating a heterostructure. A lattice system provides several
advantages. Unlike the continuum case, a localized
wavepacket can be put into oscillatory motion 
by tilting the lattice \cite{bloch,zener}
to generate Bloch oscillations \cite{superlattice_1,superlattice_2}, 
avoiding the
need to carry the electron from place to place in a trap. Furthermore, 
the wavepacket will not be excited out of is ground state by the
motion \cite{cadez_2}, avoiding a possible source of noise. By adjusting
the Rashba coupling in phase with the Bloch oscillations, we will
show how it is possible to controllably rotate the spin of an electron, 
thereby forming
a ``Bloch-Rashba rotator''. Even if the Rashba coupling can only be
varied by a small amount,
a large spin-rotation can be built up by allowing the
particle to undergo several oscillations, allowing the rotation angle
to accumulate little by little.

\section{Bloch-Rashba Hamiltonian}
We consider a one-dimensional wire lying on a
two-dimensional interface in the $x-y$ plane,
subject to a Rashba SOC governed by an electric
field perpendicular to the interface.
For convenience we will take the wire to be aligned along the $x$-direction.
In a continuum, the Rashba Hamiltonian will be given by 
$H_R = \alpha(E_z) / \hbar \ \left( {\bf \sigma} \times {\bf p} \right)_z$ 
\cite{rashba},
where $\alpha$ is the Rashba coupling, regulated by the applied electric
field $E_z$, $\sigma_j$ are the Pauli spin-operators, and
$\bf p$ is the particle's momentum.
Moving to a lattice description, the continuum Hamiltonian,
$H = p^2/2 m^\ast + H_R$, becomes a tight-binding model \cite{dey}
\begin{equation}
\Htb =  - \sum_{j} J \left[ c^\dagger_{j} c_{j+1} + \mathrm{H.c.} \right] 
 + \Jso \left[ c^{\dagger}_j \left( i \sigma_y \right) c_{j+1} + \mathrm{H.c.} \right] 
\label{lattice_ham}
\end{equation}
where $c_j^{\dagger} = \left( c^{\dagger}_{j \uparrow}, \ c^{\dagger}_{j \downarrow} \right)$, and $c_{j \sigma}^\dagger \ / \ c_{j \sigma}$ is the
creation / annihilation operator
for a fermion of spin $\sigma$ on lattice site $j$.
In this expression, $J$ represents the single-particle tunneling
between adjacent lattice sites, and $\Jso$ is the spin-orbit
tunneling produced by the Rashba SOC, whose amplitude is
proportional to $\alpha$. Clearly 
the SOC term will
induce a rotation of the electron spin around the $S_y$ axis 
when the particle moves along the lattice,
that is, the spin-rotation will be about an axis perpendicular to 
both the direction 
of motion and the direction of the electric field.

\begin{figure}
\begin{center}
\includegraphics[width=0.45\textwidth,clip=true]{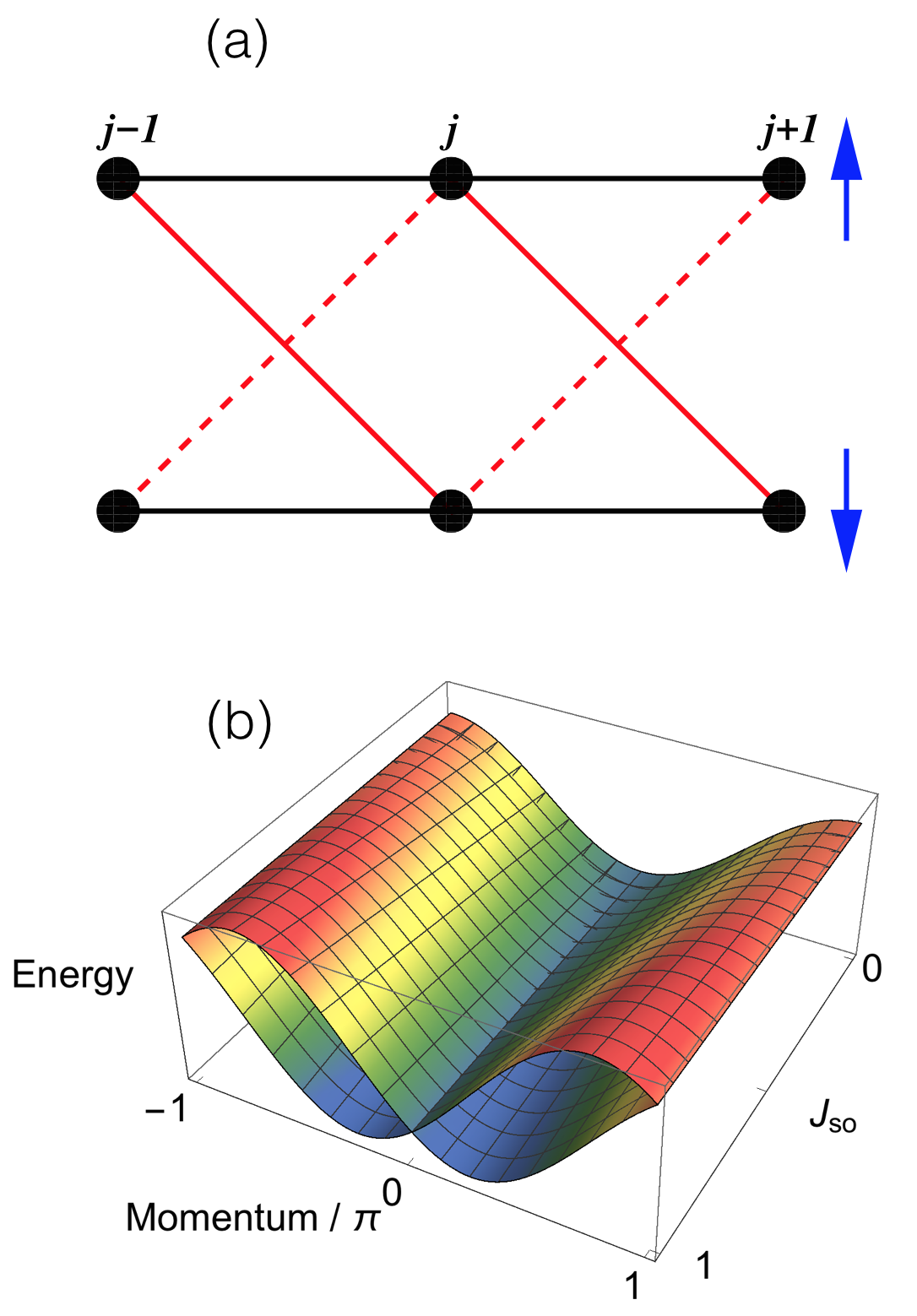}
\end{center}
\caption{(a) Creutz ladder representation of the lattice
Hamiltonian (\ref{lattice_ham}). The index $j$ labels the sites of the lattice.
Black lines along the edges of the ladder
represent standard single-particle hopping between neighboring sites which 
conserves the spin-orientation. Diagonal hopping terms (shown in red)
represent processes in which a particle hops by one lattice site
and flips its spin, which arise from the Rashba interaction.
The red dotted / solid lines have amplitudes of $- \Jso \  / \ \Jso$, 
due to the $\sigma_y$ term in $\Htb$.
(b) Dispersion relation of the lattice Hamiltonian. For $\Jso = 0$
the system exhibits the standard single-band dispersion
relation $E_k = -2 J \cos k$. As $\Jso$ is increased, the spectrum
splits into two cosinusoidal bands, displaced from
the origin by an amount proportional to the
spin-orbit coupling.}
\label{ladder}
\end{figure}

A convenient way to represent the hopping terms in Eq. \ref{lattice_ham} is 
to visualize them in terms of 
the Creutz ladder \cite{creutz}, as shown in Fig. \ref{ladder}a.
In this picture, spin-up fermions occupy sites on the top edge of the
ladder, while sites on the lower edge hold spin-down
fermions. The single-particle tunneling terms do not change the spin
of a fermion, and so they represent hopping processes along the edges
of the ladder, shown by the black lines. 
The tunneling terms governed by $\Jso$, however,
involve a spin-flip, and so are represented by the diagonal red lines 
connecting sites on the two edges of the ladder.

The dispersion relation of Hamiltonian (\ref{lattice_ham}) is
shown in Fig. \ref{ladder}b. When the
Rashba coupling vanishes ($\Jso = 0$) we recover the standard
cosinusoidal dispersion relation for a single-band
tight-binding model, each state having a two-fold
spin degeneracy. As $\Jso$ increases, the degeneracy between
spin-up and spin-down states is lifted, and the spectrum splits into 
two bands, each displaced from the origin by a momentum proportional 
to the Rashba coupling $\Jso$.

\begin{figure}
\begin{center}
\includegraphics[width=0.45\textwidth,clip=true]{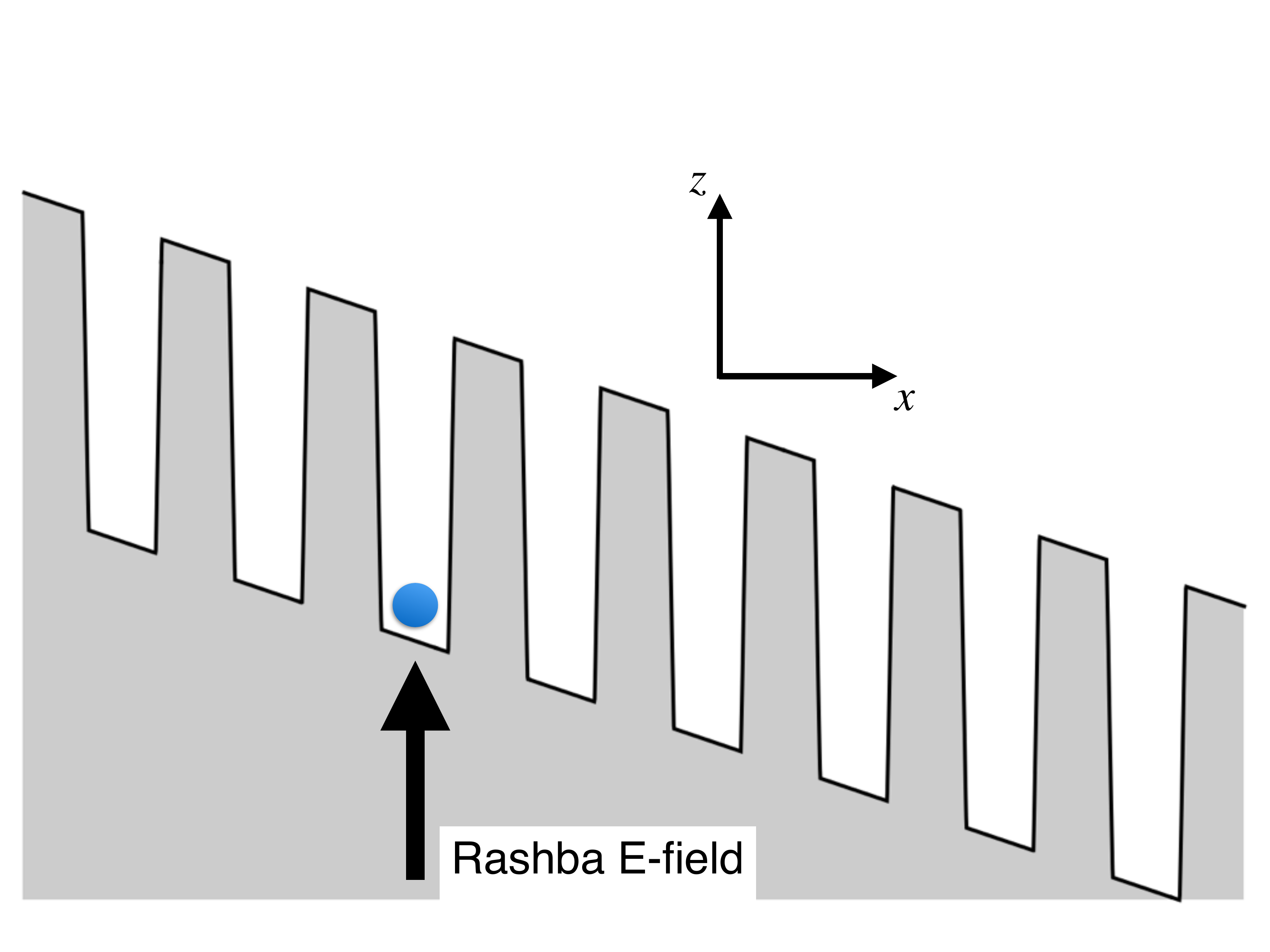}
\end{center}
\caption{Schematic form of the Bloch-Rashba rotator. A particle is
placed in a lattice which is subjected to a small tilt
causing the particle to undergo an oscillatory motion
through the lattice (Bloch oscillation). An external electric field
regulates the Rashba spin-orbit coupling, causing the spin of the
particle to rotate about an axis mutually perpendicular to its motion
and the Rashba field. We consider the particle motion to be
in the $x$-direction, while the Rashba field is aligned  with the $z$-axis;
consequently the particle spin will rotate about the
$y$-axis, in the $S_x-S_z$ plane.}
\label{schematic}
\end{figure}

Having obtained the lattice Hamiltonian (\ref{lattice_ham}), 
the next step is to introduce a tilt to the lattice potential, 
as shown in Fig. \ref{schematic}.
This is described by the Bloch-Rashba Hamiltonian
\begin{equation}
H_{\mathrm{BR}} = \Htb + V_0 \ \sum_j j n_j \ ,
\label{hamiltonian_BO}
\end{equation}
where $V_0$ is the difference in potential between neighboring
sites, and $n_j$ is the standard number operator.
Classically one would expect a particle held in 
in a tilted potential to roll down the slope
and thus accelerate uniformly to the right.
Quantum effects produced by the lattice, however, complicate
this simple picture, and the wavepacket instead undergoes
a coherent oscillation \cite{dynamics_bloch} termed Bloch oscillation, whose
frequency and amplitude depend on the lattice tilt.
If the initial wavepacket is well-localized in space, the position of
its center of mass \cite{dynamics_bloch, holthaus_bloch}
is given by the simple expression 
\begin{equation}
x(t) = 2 \left( J / V_0 \right) \ \left( 1 - \cos V_0 t  \right) \ .
\label{com}
\end{equation}

\begin{figure}
\begin{center}
\includegraphics[width=0.45\textwidth,clip=true]{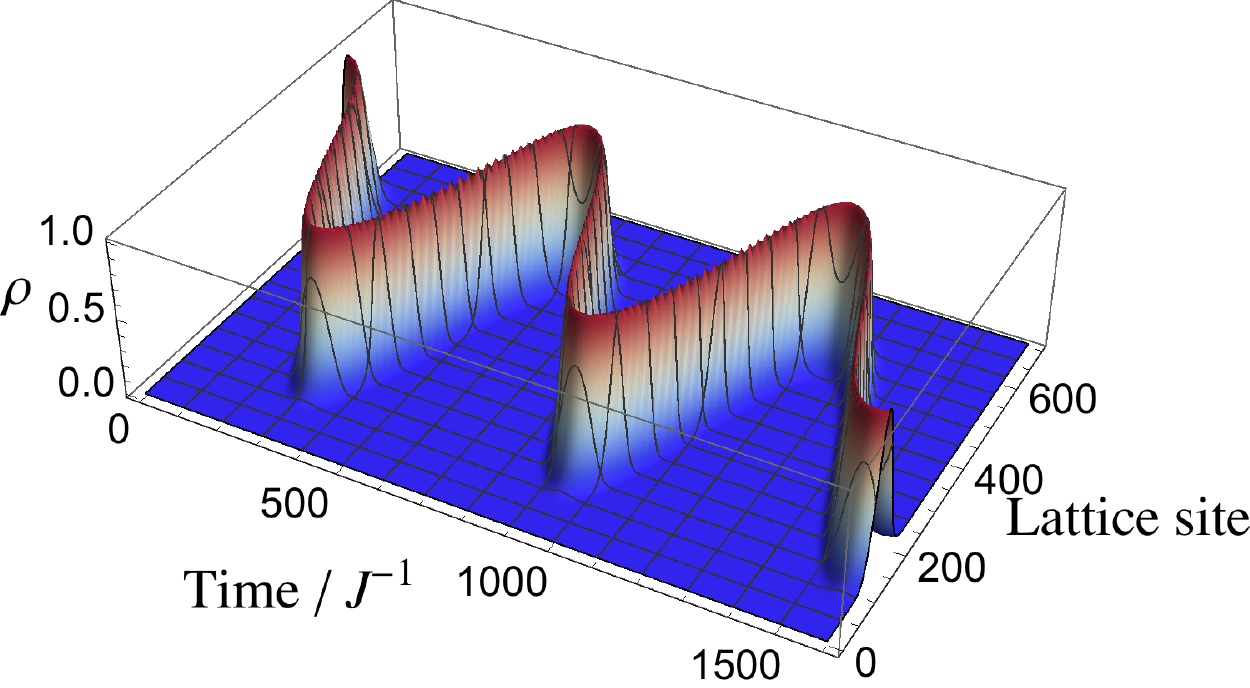}
\end{center}
\caption{A Gaussian wavepacket placed in a tilted lattice potential makes
an oscillatory motion along the lattice. The amplitude of the
oscillation and its frequency are governed by the size of the tilt, 
Eq. \ref{com}. The lattice tilt used here, $V_0 = 0.01 J$, gives a Bloch 
period of $T_B = 200 \pi/J$.}
\label{bloch_oscs}
\end{figure}

\section{Results \label{results}}

\subsection{Constant SOC}
In Fig. \ref{bloch_oscs} we show the time evolution of a Gaussian wavepacket
placed in a lattice with a small tilt of $V_0 = 0.01 J$.
The width of the Gaussian, $\sigma^2 = 1000 a^2$, 
where $a$ is the lattice spacing,
was chosen to be sufficiently small for
the wavepacket to be well-localized in space, but large enough for
it to exhibit well-defined Bloch oscillations; note that a very narrow
wavepacket would instead experience Wannier-Stark localization 
\cite{dynamics_bloch}.
The system was numerically integrated in time under the lattice Hamiltonian
(\ref{hamiltonian_BO}), and the plot displays the particle density,
$\rho(x,t) = | \psi(x,t) |^2$.
It can be seen that
the wavepacket oscillates along the lattice as expected, while retaining
its Gaussian shape, clearly displaying Bloch oscillations.

\begin{figure}
\begin{center}
\includegraphics[width=0.5\textwidth,clip=true]{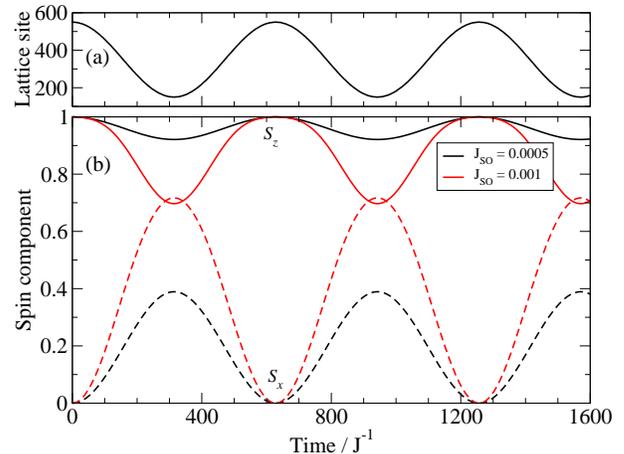}
\end{center}
\caption{(a) Bloch oscillation of the Gaussian wavepacket.
The center of mass of the wavepacket makes a sinusoidal oscillation along the
lattice, the period of which is determined by the lattice
tilt, $T_B = 2 \pi / V_0$ (see Eq. \ref{com}).
b) Solid lines denote $\sz$, the dashed lines denote $\sx$. The $x$
and $z$ components of the particle's spin oscillate sinusoidally
in time, with the same
period as the Bloch oscillation. When the SOC is increased
from $\Jso = 0.0005J$ (black lines) to $\Jso = 0.001$ (red lines), the 
amplitude of the oscillations in $\sx$ and $\sz$ correspondingly increases.}
\label{spin}
\end{figure}

In Fig. \ref{spin}a we show this motion more quantitatively,
by plotting the motion of the center of mass of the system. 
From Eq. \ref{lattice_ham}, we can see that the motion of the
particle will be associated with a rotation of its spin about
the $S_y$-axis. Thus if the particle is initialized in a spin-up
state, during its motion through the lattice, its spin will rotate
in the $S_x - S_z$ plane. In Fig. \ref{spin}b
we plot the expectation values of the spin projections $\sz$
and $\sx$  for two different values of the Rashba SOC.
We can see that in each case, $\sz$ takes an initial value of one,
as expected, and then decreases as the particle moves through the lattice.
At the same time $\sx$ increases from zero. After reaching the
extremum of its motion, the particle reverses its direction
and returns to its initial position. In this portion of its motion
the values of $\sz$ and $\sz$ return smoothly to their initial
values. The spin of the particle thus rotates periodically
in the $S_x - S_z$ plane,
with the same period as the Bloch oscillation.
In the inset of Fig. \ref{protocols} we plot the evolution
of the particle's spin in this plane.
We can note that the modulus of the spin remains approximately
constant, indicating that the spin vector is evolving
smoothly along a circle on the surface of the Bloch sphere, as required.
Close examination of this trajectory, however, reveals minor
deviations from the circle, corresponding to a periodic ``breathing''
motion of the wavepacket's width during the Bloch oscillation.
As
mentioned previously, the rotation angle acquired in the first part 
of the oscillation is exactly canceled  when the particle returns
to its original position, and so the Bloch vector only traces out a small,
retracing arc on the Bloch sphere. 
To obtain a net rotation it is 
necessary to also vary the SOC with time, and so consider a
{\em two-parameter} driving.

\subsection{Time-dependent SOC}
\begin{figure}
\begin{center}
\includegraphics[width=0.5\textwidth,clip=true]{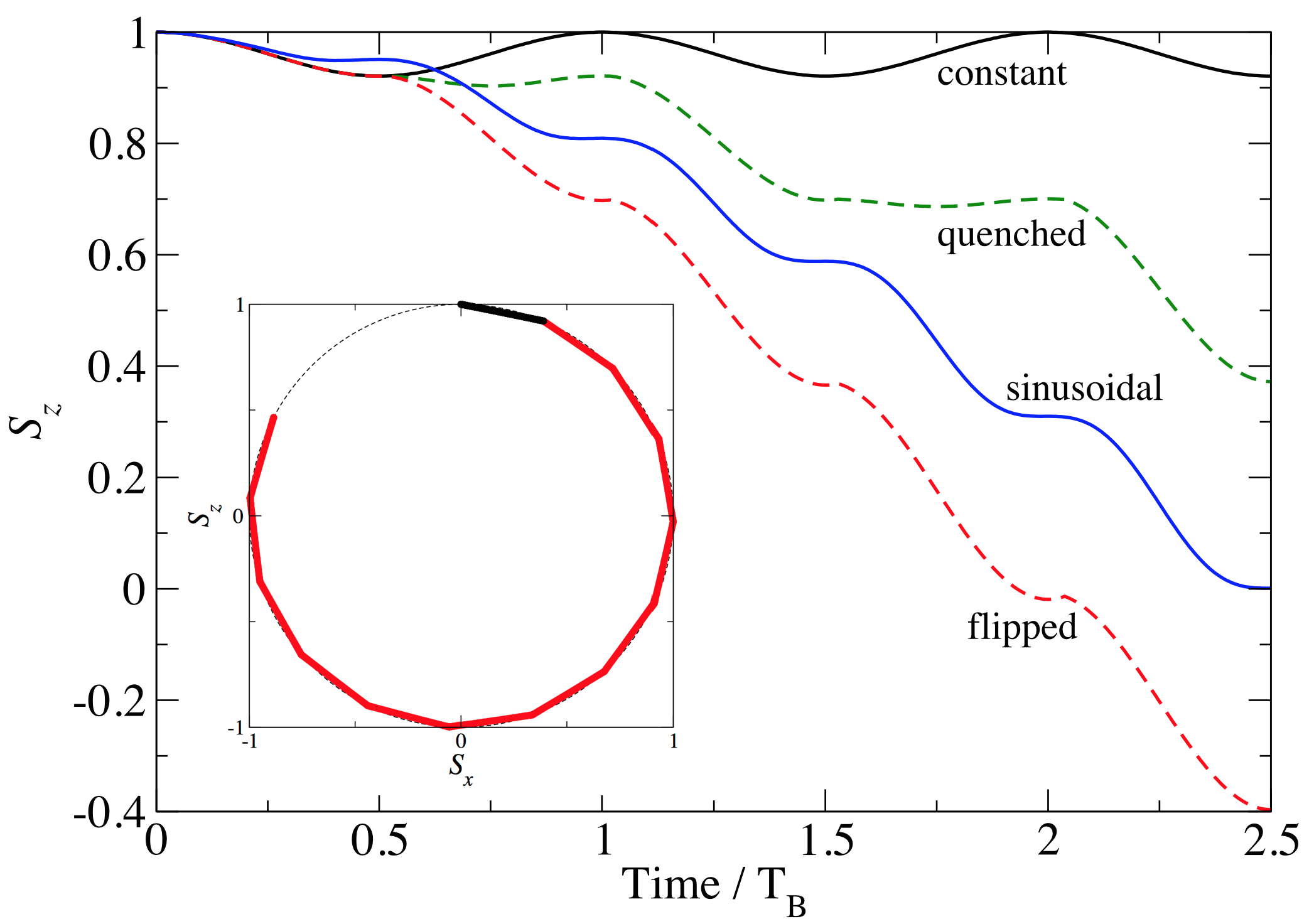}
\end{center}
\caption{The spin projections, $\sx$ and $\sz$ as a
function of time. When the Rashba tunneling, $\Jso$, is held constant,
the electron spin rotates at a constant rate while the electron
wavepacket propagates in one direction through the lattice. However,
when the wavepacket reverses to complete a cycle of Bloch
oscillation, the spin retraces its trajectory to its original configuration,
and so no spin-rotation is acquired.
When $\Jso$ is quenched to zero during the second half of the
Bloch cycle, the electron spin is frozen, and so does not
retrace its trajectory. Accordingly the rotation angle changes
in steps as the Bloch oscillation continues.
Flipping the sign of the Rashba tunneling, $\Jso \rightarrow - \Jso$
in the second
half-cycle causes the spin to continue rotating at the same rate
during the entire Bloch oscillation. $\Jso$ can also be varied
continuously, $\Jso = J_0 \sin \omega_B t$, with the same frequency
as the Bloch oscillation, to achieve this effect.
Inset: When $\Jso$ is held constant, the Bloch vector oscillates over
a small range (black symbols). By making $\Jso$ time-dependent, the
Bloch vector can now progressively step around a great circle
in the $S_x - S_z$ plane.}
\label{protocols}
\end{figure}

The simplest form of varying the SOC to produce a net spin-rotation
is for it to take two
different values \cite{kregar}: one during the outward motion of the 
Bloch oscillation, and another value while the particle returns.
In Fig. \ref{protocols}  we show the time-dependence of $\sz$
for this driving protocol, where $\Jso$ is set to zero on the return
leg. Initially the behaviour of $\sz$ follows that of
the system considered previously, but on the return leg the value
of $\sz$ is quenched. If $\Jso$ is then periodically restored and
quenched in this way, in phase with the Bloch oscillation,
the time-evolution of $\sz$ will show a staircase behavior,
Thus even if $\Jso$ is limited to a small maximum value, a large
spin-rotation can nonetheless be obtained by allowing the particle to
accumulate the rotation angle over many Bloch oscillations,
as can also be seen in the inset of the figure.

Clearly the spin-rotation will occur more rapidly if
the spin is able to continue rotating 
in the same sense on the return leg of the cycle,
rather than just being frozen.
As the Rashba SOC 
has the schematic form $H_R~=~\alpha \ \sigma_y \ p$, 
we can see that this can be done by reversing the sign
of the coupling, $\alpha \rightarrow - \alpha$, to compensate
for the reversal of the particle's momentum.
The time evolution produced by this ``flipped'' protocol
(where the sign of $\Jso$ is flipped in each half-period
of the Bloch oscillation) is also shown in Fig. \ref{protocols},
and indeed demonstrates how the rotation occurs more quickly,
the spin-rotation accumulating twice as quickly as in
the quenched protocol.

As well as using discrete values of $\Jso$, it is also possible
to vary the SOC continuously in time. In Fig. \ref{protocols} we
show the result of sinusoidally modulating $\Jso$ with the 
same period as the Bloch oscillation, 
$\Jso = J_0 \sin \left(\omega_B t \right)$.
In this case, as in the case of the ``flipped driving'', the
spin-rotation continues in the same sense in both halves of the
Bloch oscillation. As a result the rotation angle increases at a
comparable, though slower, rate to that of the case of flipped driving. 

\begin{figure*}
\begin{center}
\includegraphics[width=1.0\textwidth,clip=true]{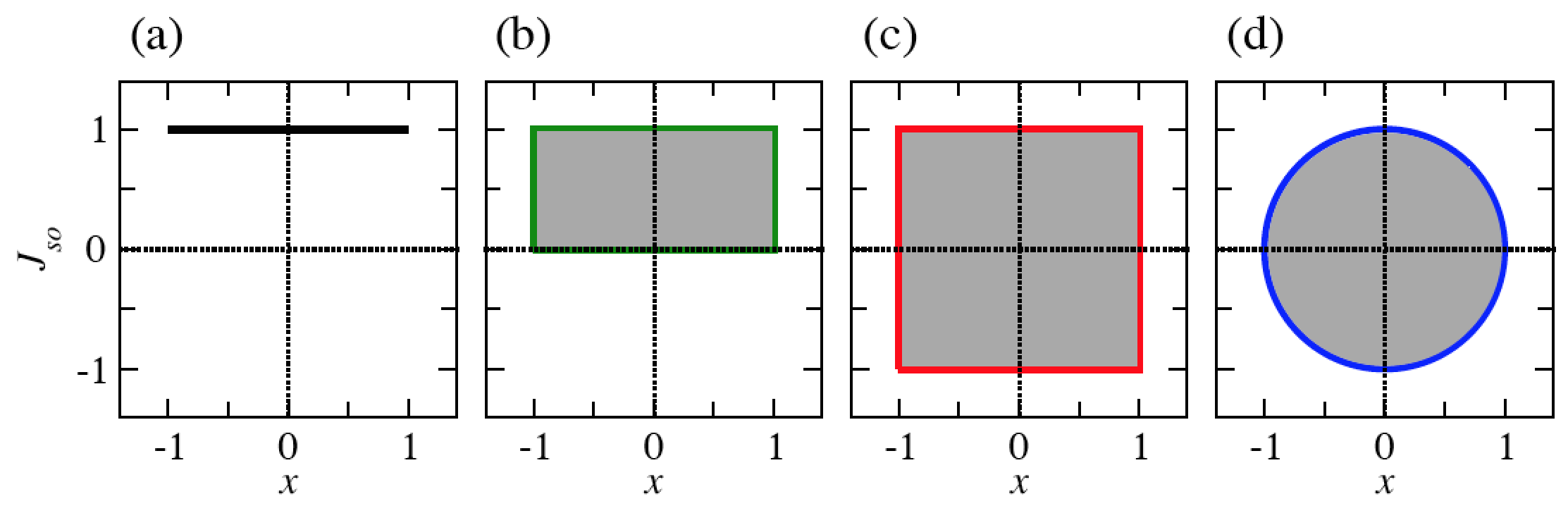}
\end{center}
\caption{Area enclosed in the parameter plane ($x - \Jso$) for the 
different protocols shown in Fig. \ref{protocols}.
The displacement $x$ is measured in units of the
amplitude of the Bloch oscillation.
(a) Constant $\Jso$. The trajectory is a straight line; as it does not
enclose an area, the net spin-rotation is zero.
(b) Quenched driving. $\Jso$ takes two values, $\Jso=1$ in the
first half-period, and $\Jso = 0$ in the second, so the trajectory
encloses a rectangle.
(c) Flipped driving. $\Jso$ again takes two values, but is now
negative ($\Jso = -1$) in the second half-period.
The area again is again rectangular, but encloses twice the
area obtained for quenched driving. As a consequence the spin rotates 
more rapidly.
(d) Sinusoidal driving. The trajectory now encloses a circle.}
\label{loop}
\end{figure*}

To compare the efficacy of the different driving protocols,
it is informative to look at the trajectory traced
out in the displacement-$\Jso$ parameter space. The net
spin-rotation achieved after one cycle of driving (one Bloch
oscillation) is proportional to the area enclosed by
this trajectory \cite{toni_john,gate_noise}. We show the four cases
that we have considered in Fig. \ref{loop}. 
When $\Jso$ is held constant, the trajectory just traces
a straight line (Fig. \ref{loop}a) which
encloses no area, and thus corresponds to
no net spin-rotation. In the quenching protocol (Fig. \ref{loop}b),
$\Jso$ takes two values and the trajectory 
traces out a rectangle. If $\Jso$ is restricted to take only
positive values, this form of driving clearly maximizes the
possible area enclosed, and so will be the most effective.
If $\Jso$ can take both positive and negative values then
the flipped driving will be the most effective,
enclosing double the area of the quenched driving protocol.
Finally, the sinusoidal driving traces out a circular trajectory
in this parameter-space. Although the rotation per cycle is less than
for flipped driving, it is of similar order.

\section{Conclusions}
We have shown how the interplay between Bloch oscillations
and the SOC can be used to create a controllable spin rotator,
that does not require an externally applied magnetic field.
In contrast to previous proposals, the electron does not
have to be transported in a moving potential well, but its
motion is instead an intrinsic property of the lattice system.
We have shown how the system can be conveniently mapped to
the Creutz ladder, and how the spin-rotation produced per
cycle can be optimized by maximizing the area enclosed by
the trajectory in the $x - \Jso$ parameter space.

In a doped InAs heterostructure, it was found that
the Rashba SOC could be enhanced by a factor of 1.5 \cite{nitta} by
applying a gate voltage of a few volts, while enhancement of
up to a factor of six could be obtained
in an InAs quantum wire \cite{liang}. 
Ferroelectric Rashba materials \cite{ferro_1,ferro_2,ferro_3} 
also hold out the prospect
of having large, electrically controllable Rashba couplings.
However, even if the Rashba coupling of a material can only be changed
by a small amount, the method described here allows
large values of spin-rotation to be achieved by letting a particle undergo
several periods of Bloch oscillation, and allowing the rotation to accumulate.
As well as applying to solid state materials,
an exciting possibility is to
use this technique to manipulate ultracold quantum gases. In such systems
an effective Rashba coupling can be engineered by dressing atomic spin states
with lasers \cite{BEC_soc}, and a lattice structure
can be imposed by  applying an optical lattice potential
\cite{bloch_soc_1,bloch_soc_2}. These systems are extremely
clean and controllable, and would provide an ideal format to investigate
this form of spin control.

In this work we have just considered a one-dimensional system, and
consequently the spin-rotation only occurs in the
$S_x - S_z$ plane. To obtain full coverage of the Bloch sphere,
it would be necessary for the particle to move in a perpendicular
direction as well. This could be achieved by applying a two-dimensional
lattice potential, in which Bloch oscillations could be induced in
the two directions. Extending the model to treat this situation,
and including the effects of noise and dissipation, are fascinating
subjects for future study. 

\acknowledgments
This work has been supported by Spain's MINECO through
grant FIS2017-84368-P. The author thanks Toni Ram{\v s}ak
for introducing him to this to this problem, and
Gloria Platero for stimulating discussions.

\bibliographystyle{aipnum4-1}
\bibliography{rashba_bib}

\end{document}